\documentclass[aps,epsfig,graphicsfloatfix,mathbbm,a4paper,nofootinbib]{revtex4}

\usepackage{amsmath,amsfonts,amssymb,graphics,graphicx,epsfig,color,times,bbm}

\begin{document}

\bibliographystyle{apsrev}
\newcommand{\spec}{\rm{spec}}
\newcommand{\id}{\rm{id}}
\newcommand{\R}{\mathbbm{R}}
\newcommand{\cB}{{\cal B}}
\newcommand{\cH}{{\cal H}}
\newcommand{\rr}{\mathbbm{R}}
\newcommand{\cT}{{\cal T}}
\newcommand{\cV}{{\cal V}}
\newcommand{\cL}{{\cal L}}
\newcommand{\E}{{\cal E}}
\newcommand{\cX}{{\cal X}}
\newcommand{\V}{{\cal V}}
\newcommand{\cc}{{\cal{C}}}
\newcommand{\ii}{\mathbbm{1}}
\newcommand{\cM}{{\mathcal M}}
\newcommand{\ra}{\rightarrow}
\newcommand{\C}{\mathbb{C}}
\newcommand{\1}{\mathbbm{1}}
\newcommand{\F}{\mathbbm{F}}
\newcommand{\h}{\frak{H}}
\newcommand{\tr}[1]{{\rm tr}\left[#1\right]}
\newcommand{\gr}[1]{\boldsymbol{#1}}
\def\>{{\rangle}}
\def\<{{\langle}}
\newcommand{\be}{\begin{equation}}
\newcommand{\ee}{\end{equation}}
\newcommand{\bea}{\begin{eqnarray}}
\newcommand{\eea}{\end{eqnarray}}
\newcommand{\ket}[1]{|#1\rangle}
\newcommand{\bra}[1]{\langle#1|}
\newcommand{\avr}[1]{\langle#1\rangle}
\newcommand{\red}[1]{{\bf \textcolor{red}{{#1}}}}
\newcommand{\D}{{\cal D}}
\newcommand{\eq}[1]{Eq.~(\ref{#1})}
\newcommand{\ineq}[1]{Ineq.~(\ref{#1})}
\newcommand{\sirsection}[1]{\section{\large \sf \textbf{#1}}}
\newcommand{\sirsubsection}[1]{\subsection{\normalsize \sf \textbf{#1}}}
\newcommand{\ack}{\subsection*{\normalsize \sf \textbf{Acknowledgements}}}
\newcommand{\front}[5]{\title{\sf \textbf{\Large #1}}
\author{#2 \vspace*{.4cm}\\
\footnotesize #3}
\date{\footnotesize \sf \begin{quote}
\hspace*{.2cm}#4 \end{quote} #5} \maketitle}
\newcommand{\eg}{\emph{e.g.}~}

\newcommand{\proofend}{\hfill\fbox\\\medskip }


\newtheorem{theorem}{Theorem}
\newtheorem{proposition}{Proposition}

\newtheorem{lemma}{Lemma}

\newtheorem{definition}{Definition}
\newtheorem{corollary}{Corollary}

\newcommand{\proof}[1]{{\it Proof.} #1 $\proofend$}

\title{\begin{center}
    {  \Large The Inverse Eigenvalue Problem\\ for Quantum Channels}
\end{center}}
\author{Michael M. Wolf$^1$, David Perez-Garcia$^2$\vspace*{8pt}}
\affiliation{$^1$ Niels Bohr Institute, 2100 Copenhagen, Denmark\\
$^2$ Dpt.~An\'alisis Matem\'atico and IMI,
Universidad Complutense de Madrid, 28040 Madrid, Spain}
\date{\today}

\begin{abstract} Given a list of complex numbers $\lambda_1,\ldots,\lambda_n$, when can it be the spectrum of a quantum channel, i.e., a completely positive trace preserving map? We provide an explicit solution for the $n=4$ case and show that in general the characterization of the non-zero part of the spectrum can essentially be given in terms of its classical counterpart---the non-zero spectrum of a stochastic matrix. A detailed comparison between the classical and quantum case is given. We discuss applications of our findings in the analysis of time-series and correlation functions and provide a general characterization of the peripheral spectrum, i.e., the set of eigenvalues of modulus one. We show that while the peripheral eigen-system has the same structure for all Schwarz maps, the constraints imposed on the rest of the spectrum change immediately if one departs from complete positivity.
\end{abstract}

\maketitle

\tableofcontents


\section{Introduction and outline}
Quantum channels provide general input-output relations for quantum mechanical evolutions.  If the input space equals the output space, we can assign a spectrum to each quantum channel. The spectral properties play a role wherever successive powers of the channel are applied. These may describe discrete time-evolution or they may appear in particular representations of quantum spin chains. Apart from the most basic facts which coincide with their `classical' counterparts for stochastic matrices, not much is known about the spectral properties of completely positive maps. The present work aims at a deeper analysis, beyond Perron-Frobenius theory, and provides necessary, and in some cases also sufficient, conditions for a list of numbers to be the spectrum of a quantum channel. That is, we address the `inverse eigenvalue problem' for completely positive maps---occasionally being content with plain (not complete) positivity.

After having collected some basic results in Sec.\ref{sec:basicprop}, Sec.\ref{sec:qubitmapspec} provides a rather exhaustive analysis of qubit-maps---a complete characterization of the spectra of (completely) positive qubit maps is given. In Sec.\ref{sec:nonzerospec0} the non-zero part of the spectrum of general (completely) positive maps is investigated and close analogies to the classical case of stochastic matrices are found.
Sec.\ref{sec:peripheralspec0} provides a detailed study of the \emph{peripheral spectrum} (i.e., eigenvalues whose modulus equals the spectral radius) and of the corresponding eigenvectors. Before outlining some applications in the analysis of time-series and correlation functions in Sec.\ref{sec:appltimecorr} we compare the quantum and the classical world, i.e., the spectral properties of quantum channels and those of stochastic matrices in Sec.\ref{sec:clasVSquant}.

\section{Basic spectral properties of quantum channels}\label{sec:basicprop}
We begin with fixing some notation and recalling basic results. Let $T:\cM_d(\C)\ra\cM_d(\C)$ be a linear map on the space of complex valued $d\times d$ matrices. $T$ is said to be \emph{positive} if it maps the cone of positive semidefinite matrices into itself. If $T$ is positive, then it is in particular Hermiticity preserving, i.e., $T(A)^\dagger=T(A^\dagger)$ for all $A\in\cM_d(\C)$. The adjoint map $T^*$ is defined by imposing $\tr{A T^*(B)}=\tr{T(A)B}$ for all $A,B\in\cM_d(\C)$. $T$ is trace-preserving iff $T^*$ is unital, meaning that $T^*(\1)=\1$. A linear map on $\cM_d(\C)$ is \emph{completely positive} iff $T\otimes\id_d$ is positive. A completely positive and trace-preserving linear map is referred to as \emph{quantum channel}.

As a vector space $\cM_d(\C)$ is isomorphic to $\C^{d^2}$ so that $T$ can be represented as a $d^2\times d^2$ matrix for which we will write $\hat{T}$. Fixing a Hilbert-Schmidt orthonormal basis in $\cM_d(\C)$, i.e., a collection of $d^2$ operators $\{E_i\in\cM_d(\C)\}$ with $\tr{E_i^\dagger E_j}=\delta_{ij}$, we get $\hat{T}_{ij}=\tr{E_i^\dagger T(E_j)}$. If $T$ is Hermiticity preserving, then $T^*$ is represented by the adjoint matrix $\hat{T}^\dagger$.

When analyzing the spectrum of a map we distinguish between the spectrum as a set, where each element only appears once, and as a \emph{multiset} (unordered $n$-tuple), where elements appear according to their algebraic multiplicity. More precisely, the \emph{spectral set} of $T$ is the set of complex numbers $\lambda$ for which $T-\lambda\id$  is not invertible. The spectrum (as a multiset) is the list of roots of the characteristic polynomial $\det[\hat{T}-\lambda\1]$ where each root appears with a multiplicity according to its degeneracy. We will write $\spec(T)$ if we refer to the multiset and $\{\spec(T)\}$ if the spectral set is meant.

From Hermitian conjugation of the eigenvalue equation $T(X)=\lambda X$ we see that the eigenvalues of Hermiticity preserving maps are either real or come in complex conjugate pairs. This holds even if we take multiplicities into account. Since $\spec(T)=\spec(T^*)$, every trace-preserving map has an eigenvalue one. If the map is positive in addition, then its spectral radius is one, i.e., $\max\{|\lambda|\;\big|\lambda\in\spec(T)\}=1$.

A positive map is  \emph{irreducible} \cite{EH78} iff the spectral radius is a non-degenerate eigenvalue and the corresponding eigen-vector is positive definite. If in addition there is no other eigenvalue whose modulus matches the spectral radius then the map is called \emph{primitive} \cite{SPWC09}.

 The following shows that a characterization of the spectra of (completely) positive maps is essentially independent of whether or not the maps are required to be trace-preserving:

\begin{proposition}[Similarity with trace-preserving maps]\label{prop:simspectracepr}
Let $T:\cM_d(\C)\ra\cM_d(\C)$ be a (completely) positive map with spectral radius $\varrho>0$. There is a trace-preserving (completely) positive map $T'$ on $\cM_d(\C)$ such that $\spec(T)=\varrho\;\spec(T')$. \end{proposition}
\proof{ Define a map $T_\epsilon:X\mapsto T(X)+\epsilon\tr{X}\1$ and denote its spectral radius by $\varrho_\epsilon$. Following \cite{EH78} $T_\epsilon$ is irreducible for all $\epsilon>0$ and for $\epsilon\ra 0$ we have that $\spec(T_\epsilon)\ra\spec(T)$ continuously. Irreducibility implies the existence of a positive definite $P=T_\epsilon(P)\varrho_\epsilon^{-1}>0$. Hence the map $T_\epsilon'(X):=\varrho_\epsilon^{-1}P^{-1/2}T_\epsilon(P^{1/2}X P^{1/2})P^{-1/2}$ satisfies $\spec(T_\epsilon)=\spec(T_\epsilon')\varrho_\epsilon$ and is (completely) positive and  unital. Since these maps form a compact set, the limit $T'=\lim_{\epsilon\ra 0}(T_\epsilon'^*)$ exists and has all desired properties.
}

In the following we will consider trace-preserving maps only.

\section{The $n=4$ case -- qubit channels}\label{sec:qubitmapspec}

We continue the discussion with completely positive qubit maps  $T:\cM_2(\C)\rightarrow \cM_2(\C)$  which allow for an exhaustive analysis. Some of the results derived in this section will be the basis for the general characterization of the non-zero part of the spectrum in Sec.\ref{sec:nonzeroset}. We will first recall and derive some preparatory results and then provide an explicit solution of the inverse eigenvalue problem in Thm.\ref{thm:qubitsolution}.

Choosing normalized Pauli matrices as operator basis  (with $\sigma_0=\1$) we can represent $T$ as a $4\times 4$ matrix $\hat{T}_{ij}:=\tr{\sigma_i T(\sigma_j)}/2$. If $T$ is Hermiticity preserving, then  $\hat{T}$ is real, and if $T$ is trace preserving, then $(T_{1,j})=(1,0,0,0)$. Since the maps of our interest satisfy both, they can be represented as
\be \hat{T}=\left(\begin{array}{cc}1 & 0\\  v & \Delta
\end{array}\right),\label{eq:2channelDelta} \ee
where $\Delta$ is a real $3\times 3$ matrix and $v\in\mathbb{R}^3$.
The spectrum of $T$ is thus the union \be\label{eq:spec2Delta}\spec(T)=\{1\}\cup\spec(\Delta).\ee
The  Jamiolkowski state corresponding to $\hat{T}$ is given by $\tau=\frac14\sum_{ij}\hat{T}_{ij}\sigma_i\otimes\sigma_j^T$ so that $v$ describes its reduced density operator and $\Delta$ its correlations. The (generally cumbersome) conditions for $\hat{T}$ to correspond to a completely positive map have to be read off from $\tau\geq 0$ (cf.~\cite{Ruskai2002}). The parametrization allows, however, for a nice geometric interpretation: parameterizing a density operator via $\rho=(\1+\sum_{k=1}^3x_k\sigma_k)/2$, i.e., in terms of a vector $x\in\mathbb{R}^3$ within the Bloch ball $||x||_2\leq 1 $, the action of $T$ as parametrized in Eq.(\ref{eq:2channelDelta}) is a simple affine transformation
\be x\mapsto v+\Delta x.\label{eq:affinetrafo1}\ee
From here conditions for $T$ to be positive are readily derived as the vector has to stay within the Bloch ball. As the center of the Bloch ball corresponds to the maximally mixed state, $v=0$ holds iff $T$ is unital. In this case positivity is equivalent to $||\Delta||_\infty\leq 1$.
The following is a useful proposition which allows us to restrict ourselves to the unital case:
\begin{proposition}[Reduction to unital maps]\label{prop:redt2unital} Let $T:\cM_2(\C)\ra\cM_2(\C)$ be a trace-preserving linear map with $\Delta_{ij}:=\tr{\sigma_i T(\sigma_j)}/2$ the lower-right submatrix (i.e., $i,j=1,2,3$) of the matrix representation $\hat{T}$. Then the  map defined by $\hat{T}':=1\oplus\Delta$ is unital,  trace-preserving and has the same spectrum as $T$. Moreover, it is (completely) positive if $T$ is.
\end{proposition}
\proof{By construction $T'$ is trace-preserving and unital and has the same spectrum as $T$, so it remains to show that (complete) positivity is preserved when going from $T$ to $T'$. To this end we use that $D:={\rm diag}(1,-1,-1,-1)$ is the matrix representation of time-reversal, i.e., matrix transposition in some basis. A look at the Choi matrix reveals that the map $D\hat{T}D$ is completely positive if $\hat{T}$ is. Similarly, if $T$ is merely positive, then $D\hat{T}D$ is positive as well, since it is a concatenation of positive maps. Consequently, the convex combination $(\hat{T}+D\hat{T}D)/2=\hat{T}'$ inherits the property of being (completely) positive from $T$.\footnote{Note that in higher dimensions the strategy of the proof fails: the higher-dimensional analogue of the above $D$ maps $\rho\mapsto 2\tr{\rho}\1/d-\rho$, which is no longer a positive map if $d>2$. Similarly, $D\hat{T}D$ then can fail to be completely positive even if $T$ is so.}
}

For a  qubit map of the form in Eq.(\ref{eq:2channelDelta}) with $v=0$ complete positivity can be expressed in terms of the singular values of $\Delta$ and its determinant. In order to understand this,  suppose that one acts with a unitary before and another one after applying the map $T$ so that the overall action is $\rho\mapsto U_2T(U_1\rho U_1^\dagger)U_2^\dagger$. The   matrix representing this concatenation is then given by the product $(1\oplus O_2)\hat{T}(1\oplus O_1)$, where $O_i\in SO(3)$ are real rotations of the Bloch sphere. This reflects the two-to-one group homomorphism $SU(2)\rightarrow SO(3)$. Hence, we can use this  to diagonalize $\Delta\rightarrow{\rm diag}(s_1,s_2,s_3)$ without changing the complete positivity property. Moreover, we can choose the $O_i$'s such that the components of $s$ follow a particular order and satisfy $s_1\geq s_2\geq |s_3|$.  Then a generally necessary and for unital maps also sufficient condition for complete positivity is that \cite{Verstraete2002}
\be s_1\leq 1\quad\mbox{and}\quad s_1+s_2\leq 1+s_3.\label{eq:necpos2}
\ee
These  conditions can, alternatively, be summarized by ``$s\in\cT$" \cite{Ruskai2002} where $\cT\subset\R^3$ is the tetrahedron whose corners have coordinates with components $s_i=\pm 1$ satisfying $s_1s_2s_3=1$. Note that $s\in\cT$ is independent of the ordering and sign pattern attained from a particular choice of the $O_i\in SO(3)$.

Since our interest lies, following Eq.(\ref{eq:spec2Delta}), in the eigenvalues of $\Delta$ but the constraints are essentially given in terms of its singular values we need to relate singular values and eigenvalues. This is the content of a classic result of Weyl~\cite{Weyl49} and Horn~\cite{Horn54} of which we need a real version provided in \cite{LiMa01}. Applied to our context this gives:

\begin{lemma}\label{lem:slambda} Given $\lambda\in\C^3$ and $s\in\R^3$, there is a $\Delta\in\cM_3(\R)$ with $\spec{(\Delta)}=\lambda$ and $O_1\Delta O_2={\rm diag}(s_1,s_2,s_3)$ for some $O_i\in SO(3)$ iff the following three requirements are fulfilled:
\begin{enumerate}
    \item  $\lambda$ is, as a multiset, invariant under complex conjugation, i.e., either $\lambda\in\R^3$  or there is a single real $\lambda_i$ and a complex conjugate pair.
\item $\prod_{i=1}^3 s_i = \prod_{i=1}^3 \lambda_i$.
\item $|s_1^\downarrow s_2^\downarrow|\geq |\lambda_1^\downarrow\lambda_2^\downarrow|$ and $|s_1^\downarrow|\geq|\lambda_1^\downarrow|$ where $^\downarrow$ refers to a (re-)ordering with decreasingly ordered absolute values, i.e., $|s_i^\downarrow|\geq|s_{i+1}^\downarrow|$ and similarly $|\lambda_i^\downarrow|\geq|\lambda_{i+1}^\downarrow|$.
\end{enumerate}
\end{lemma}
\proof{
\emph{Necessity:} 1. follows from $\Delta$ being real, 2. follows from both sides being equal to $\det(\Delta)$, and 3. follows from Weyl's theorem~\cite{Weyl49} since the $|s_i|$ are the singular values of $\Delta$.

\emph{Sufficiency:} Following \cite{LiMa01} the conditions 1.-3. guarantee the existence of a matrix $M\in\cM_3(\R)$ with $\spec(M)=\lambda$ and singular values $|s_i|$. The latter implies that there are $\tilde{O}_i\in SO(3)$ s.t. $\tilde{O}_1M\tilde{O}_2={\rm diag}(\tilde{s}_1,\tilde{s}_2,\tilde{s}_3)$ where $|\tilde{s}_i|=|s_i|$. Moreover, since $\prod_i s_i=\prod_i\tilde{s}_i$ (exploiting condition 2.) we have that $s$ and $\tilde{s}$ differ either by no, or by two signs. In both cases we can set $\Delta:=M$. If two signs are different we simply multiply one of the $\tilde{O}_i$'s by a diagonal matrix like ${\rm diag}(1,-1,-1)$ or a permutation thereof in order to obtain the $O_i$'s.  }

Now we are prepared for solving the inverse eigenvalue problem for qubit channels:

\begin{theorem}[Spectra of qubit channels]\label{thm:qubitsolution} Given a multiset $\Lambda\in\C^4$ the following statements are equivalent:
\begin{enumerate}
    \item There is a trace-preserving completely positive linear map $T:\cM_2(\C)\ra\cM_2(\C)$ such that $\spec(T)=\Lambda$.
	\item There is a unital and trace-preserving completely positive linear map $T:\cM_2(\C)\ra\cM_2(\C)$ such that $\spec(T)=\Lambda$.
	\item $\Lambda=\{1\}\cup\lambda$ where $\lambda\in\C^3$ is, as a  multiset, closed under complex conjugation. Furthermore, if we define $s\in\R^3$ by $s_i:=\lambda_i$ if $\lambda_i\in\mathbb{R}$ and $s_i:=|\lambda_i|$ otherwise, then (with $\cT$ the tetrahedron defined below Eq.(\ref{eq:necpos2}))\be s\in\cT\label{eq:main2speccond}.\ee
\end{enumerate}
\end{theorem}
\proof{ The equivalence 1.$\Leftrightarrow$2. is a direct consequence of Prop.\ref{prop:redt2unital}.

3.$\Rightarrow$2.: by Lemma~\ref{lem:slambda} there exists a $\Delta\in\cM_3(\R)$ with $\spec(\Delta)=\lambda$ and $O_1\Delta O_2={\rm diag}(s_1,s_2,s_3)$ for some $O_i\in SO(3)$. The unital and trace-preserving map defined via $\hat{T}=1\oplus\Delta$ has thus $\spec(T)=\{1\}\cup\lambda=\Lambda$ and is completely positive due to Eq.(\ref{eq:main2speccond}).

2.$\Rightarrow$3.:
as discussed above, the matrix representation of the channel has the form $\hat{T}=1\oplus\Delta$ and $\spec(T)=\{1\}\cup\lambda =\Lambda$ with $\lambda=\spec(\Delta)$. Our aim is to argue that ${\rm diag}(1,s_1,s_2,s_3)$ represents a valid quantum channels as well so that Eq.(\ref{eq:main2speccond}) has to be fulfilled. Making use of the real Schur decomposition we can w.l.o.g. assume that $\Delta$ has one of the following forms
\be\label{eq:2deltas} \left(\begin{array}{ccc} \lambda_1 & * & * \\ &  \lambda_2 & * \\ & & \lambda_3
\end{array}\right),\qquad  \left(\begin{array}{ccc} \lambda_1 & * & * \\ &  a & b \\ & -b & a
\end{array}\right),\ee
where $*$ denotes arbitrary entries and the `real form' on the left is such that $\lambda_i\in\R$ and the `complex form' on the right is such that $\lambda_1\in\R$ and $\lambda_{2,3}=a\pm i b$ with $a,b\in\R$.

Let us consider the real form first. Define a new matrix $\Delta'\in\cM_3(\R)$ via the convex combination
\be \Delta':=\frac1{|S|}\sum_{D_i\in S} D_i\Lambda D_i,\label{eq:convcombrev2}\ee
where $S$ is the set of four diagonal matrices $D_i$ which have diagonal entries $\pm 1$ and satisfy $\det(D_i)=1$ so that $\Delta'={\rm diag}(\lambda_1,\lambda_2,\lambda_3)$. Note that $\hat{T}':=1\oplus\Delta'$ represents a valid quantum channel since Eq.(\ref{eq:convcombrev2}) corresponds to a convex combination of channels which merely differ by unitary conjugations (represented by $1\oplus D_i$). Therefore $\lambda\in\cT$ and since in the real case $s=\lambda$ Eq.(\ref{eq:main2speccond}) follows.

Consider now the complex form. Again we can get rid of the $*$-entries by convex combination, if we choose $S=\{\1,{\rm diag}(1,-1,-1)\}$ this time. Since the two-by-two block matrix in the lower right corner has singular values $|\lambda_2|, |\lambda_3|$ (which equal $s_2$ and $s_3$) there are special orthogonal matrices $O_1,O_2$ such that $O_1\Delta'O_2={\rm diag}(s_1,s_2,s_3)$. Consequently, $s\in\cT$ because  ${\rm diag}(1,s_1,s_2,s_3)$ represents a valid quantum channel since it can be obtained from $T$ by convex combinations and concatenating with unitary conjugations.
}

A counterpart of this theorem for positive maps can easily be derived:
\begin{theorem}[Spectra of positive qubit maps]. Let $T:\cM_2(\C)\ra\cM_2(\C)$ be a positive and trace-preserving linear map. Then the multiset $\Lambda:=\spec(T)$ satisfies
\begin{enumerate}
    \item $1\in\Lambda$ and $\Lambda$ is invariant under complex conjugation,
	\item $|\lambda|\leq 1$ for all $\lambda\in\Lambda$.
\end{enumerate}
Conversely, if a multiset $\Lambda$ of four complex numbers satisfies these two conditions, then there is a unital and trace-preserving positive linear map $T$ on $\cM_2(\C)$ such that $\spec(T)=\Lambda$.
\end{theorem}
\proof{
As discussed before, conditions 1. and 2. are necessary for positive trace-preserving linear maps in any dimension. For the converse we construct a map via $\hat{T}:=1\oplus\Delta$ with $\Delta$ of the form in Eq.(\ref{eq:2deltas}), depending on whether or not $\Lambda=\{1,\lambda_1,\lambda_2,\lambda_3\}$ contains a complex conjugate pair, and $*=0$ in both cases. By construction we have $\spec(T)=\Lambda$ and $T$ is trace-preserving and unital. Moreover, it is positive since $||\Delta||_\infty\leq 1$ guarantees that $T$ maps the Bloch sphere into itself. }

\section{The non-zero part of the spectrum}\label{sec:nonzerospec0}
\subsection{The spectral set}\label{sec:nonzeroset}

In this subsection we will regard the spectrum as a set. That is, $\{\spec(T)\}$ is a set of complex numbers which contains each element only once so that degeneracies are not taken into account. This considerably simplifies the inverse eigenvalue problem if we allow for an additional null space:
\begin{theorem}[Spectral sets without zero]\label{thm:specsets} Let $\Lambda=\{\lambda_1,\ldots,\lambda_N\}$ be a set of non-zero complex numbers such that
\begin{enumerate}
    \item it is closed under complex conjugation: $\Lambda=\bar{\Lambda}$,
\item it contains one: $1\in\Lambda$, and
\item all elements are contained in the unit disc: $|\lambda|\leq 1,$ $\forall\lambda\in\Lambda$.
\end{enumerate}
Then there exists a completely positive and trace-preserving linear map $T:\cM_d(\C)\ra\cM_d(\C)$ with $d\leq 2 \max\{N-1,1\}$ such that
\be \{\spec(T)\}\backslash \{0\}=\Lambda.\label{eq:nonzerospecset}\ee Conversely, if $\Lambda$ is the spectral set of a positive and trace-preserving linear map, then conditions 1.-3. have to be fulfilled.
\end{theorem}
\proof{ For each $\lambda_j\in\Lambda$ consider the multiset $\Lambda_j:=\{1,1,\lambda_j,\bar{\lambda}_j\}$. According to Thm.\ref{thm:qubitsolution} there is a quantum channel $T_j:\cB(\cH_j)\ra\cB(\cH_j)$, where $\cH_j\simeq \C^2$,  such that $\spec(T_j)=\Lambda_j$. Defining a Hilbert space $\cH:=\bigoplus_j \cH_j$ and isometries $V_j:\cH\ra\cH_j$ we can construct a quantum channel on $\cB(\cH)$ via \be\label{eq:Tconstspecset} T(\rho):=\sum_j V_j^\dagger T_j\Big(V_j\rho V_j^\dagger\Big)V_j.\ee
By construction $\lambda_j\in\spec(T) $ for all $\lambda_j\in\Lambda$. Moreover, since $T\big(|\psi\>\<\phi|\big)=0$ for all $\psi\in\cH_k$, $\phi\in\cH_l$ with $k\neq l$, the kernel completes the spectrum of $T$ such that indeed
\be \{\spec(T)\}=\Lambda\cup\{0\}.\ee
What is $d=\dim\cH$? Assume that $N_r$ and $N_c$ are the numbers of real elements different from $ 1$ and complex conjugate pairs in $\Lambda$ respectively, i.e., $N=N_r+2N_c+1$. Then our construction requires $d=2(N_r+N_c)\leq 2N-2$ degrees of freedom if $N\geq 2$ and $d=2$ if $\Lambda=\{1\}$.

The converse statement in the theorem is just a restatement of the general properties discussed in Sec.\ref{sec:basicprop}.
}

\subsection{Moments and the non-zero part of the spectrum}\label{Sec:momentsANDnonz}
From now on we consider the spectrum again as a multiset. That is, every eigenvalue appears according to its algebraic multipliticity.
A central notion in the following discussion of the non-zero part of the spectrum is that of \emph{moments}. Given a multiset of complex numbers $\Lambda=\{\lambda_1,\ldots,\lambda_n\}$ the $k$'th moment of $\Lambda$ is defined via
\be\label{eq:LambdaMoment} \mu_k(\Lambda):=\sum_{j=1}^n\lambda_j^k,\quad k\in\mathbb{N}.\ee
The collection of moments determines the non-zero part of $\Lambda$ and viceversa. Obviously, all moments are real if $\Lambda=\bar{\Lambda}$. Conversely, positivity of all moments $\mu_k(\Lambda)\geq 0$, $k\in\mathbb{N}$ implies $\Lambda=\bar{\Lambda}$. Positivity  also guarantees that $\Lambda$ contains a positive element $\varrho\in\Lambda$ for which $\forall\lambda\in\Lambda:|\lambda|\leq\varrho$ holds \cite{Fri09}.

Note that if $\Lambda$ is the spectrum of some linear map with matrix representation $\hat{T}$, then $\mu_k(\Lambda)=\tr{\hat{T}^k}$. If $T$ is a linear map on matrices, this can be rewritten as follows:
\begin{lemma}[Moments of maps on matrices]\label{lem:moments1} Let $T:\cM_d(\C)\ra\cM_d(\C)$
be a linear map, then \be \label{eq:momentsChoi}\tr{\hat{T}^k}=\;\<\Omega\big|\big(T^k\otimes\id\big)\big(|\Omega\>\<\Omega|\big)\big|\Omega\>,\ee where $|\Omega\>:=\sum_{i=1}^d|ii\>$. Moreover, if $T$ is completely positive and has Kraus operators $\{K_j\}$ then
\be \tr{\hat{T}^k} =\sum_{j_1,\ldots,j_k} \Big|\tr{K_{j_1}\cdots K_{j_k}}\Big|^2.\label{eq:momentsKraus}\ee\end{lemma}
\proof{ Both statements are elementary observations when using matrix units $|i\>\<j|$ as basis elements for computing the trace in Hilbert-Schmidt space as $$\tr{\hat{T}^k}=\sum_{i,j=1}^d \<i|T^k\big(|i\>\<j|\big)|j\>.$$}

This leads us to the main theorem of this section:
\begin{theorem}[Non-zero part of the spectrum]\label{thm:mainIEP}
Let $T:\cM_d(\C)\ra\cM_d(\C)$ be a trace-preserving and completely positive linear map. Then $\mu_k\big(\spec(T)\big)\geq 0$ for all $k\in\mathbb{N}$.

Conversely, consider a multiset  $\Lambda=\{\lambda_1,\ldots,\lambda_n\}$ of non-zero complex numbers such that $\max\{|\lambda|\;\big|\;\lambda\in\Lambda\}=1$ is attained for a unique element $\lambda_p\in\Lambda$. If for all $k,m\in\mathbb{N}$ \bea \mu_k\big(\Lambda\big)&\geq& 0,\quad\mbox{and}\label{eq:mucond1}\\ \label{eq:mucond2} \mu_m\big(\Lambda\big) &>& 0\quad \mbox{implies}\quad \mu_{km}\big(\Lambda\big)>0,\eea
then there exists a primitive trace-preserving completely positive linear map $T$ such that $\Lambda$ is the non-zero part of the spectrum of $T$, i.e.,  $\spec(T)\backslash\{0\}=\Lambda$.
\end{theorem}
\proof{ Positivity of the moments follows from Lemma \ref{lem:moments1} and the fact that $T$ is assumed to be completely positive. For the converse statement we exploit a highly non-trivial result of \cite{BoHa91}. This states that Eqs.(\ref{eq:mucond1},\ref{eq:mucond2}) together with the uniqueness of $\lambda_p$ are necessary and sufficient for the existence of some $N\in\mathbb{N}$ and a primitive non-negative matrix $M\in\cM_N(\R_+)$ with $\spec(M)\backslash\{0\}=\Lambda$. From this one can easily construct a stochastic matrix with the same spectrum: denote by  $\<L|=\<L|M$ the left Perron eigenvector and define a diagonal matrix $X$ with entries $X_{ii}=\<L|i\>$ for a set of orthonormal basis vectors $\{|i\>\}$. Then $S:=XMX^{-1}$ is a primitive stochastic matrix which  has the same spectrum as $M$. Now define a map $T:\cM_N(\C)\ra\cM_N(\C)$ via
\be \label{eq:TfromStoch} T(\rho):=\sum_{i,j=1}^N S_{ij}\;\<j|\rho|j\>\;|i\>\<i|.\ee
If $S|p\>=\lambda |p\>$, then $T(\rho)=\lambda\rho$ for $\rho=\sum_i p_i|i\>\<i|$ and thus $\spec{(T)}\supseteq\Lambda$, i.e., the $N$ eigenvalues of $S$ (taking algebraic multiplicities into account) appear in the spectrum of $T$. Moreover, $T\big(|i\>\<j|\big)=0$ for all $i\neq j$ gives rise to a kernel of dimension $N(N-1)$ which thus completes the spectrum so that indeed $\Lambda$ is the entire non-zero part of the spectrum of $T$. By construction $T$ is trace-preserving and completely positive. In fact, it is an \emph{entanglement breaking} quantum channel. Primitivity of $T$ follows from primitivity of $S$: since $S$ has an entrywise positive fixed point, $T$ has a positive definite fixed point. Together with the fact that $T$ has a unique eigenvalue of magnitude one this implies primitivity~\cite{SPWC09}.
}

Some remarks on the above theorem are in order:
\begin{enumerate}
	\item \emph{Positivity does not imply positive moments.} Take any positive map $T$ on $\cM_d(\C)$ which is not completely positive. Then there is a $\psi\in\C^{d^2}$ such that $\<\psi|(T\otimes\id)(|\Omega\>\<\Omega|)|\psi\>< 0$. We can always write $|\psi\>=(Y\otimes\1)|\Omega\>$ for some $Y\in\cM_d(\C)$. Using Eq.(\ref{eq:momentsChoi})  the map $T'(\rho):= Y^\dagger T(\rho)Y$ will then be such that $\tr{\hat{T}'}<0$ although $T'$ is positive. The same holds for $n$-positive maps with $n<d$.
	\item \emph{The required number of zeros.} It was shown in \cite{BoHa91} that even for the case $n=4$ there cannot be an upper bound on the number $N-n$ of zeros which one has to add in order realize a given $\Lambda$ (which satisfies all necessary requirements) as the non-zero part of the spectrum of some $N\times N$ non-negative matrix $M$. A priori, this does not imply that such a bound does not exist when the realization is in terms of a quantum channel. In fact, the inequality in Eq.(\ref{eq:JLL}) from which the necessity of a large number of zeros is usually derived in the classical case, fails to hold in the quantum case. And indeed, the additional freedom provided by quantum mechanics often allows for much more efficient realizations --- see \cite{WoPe09} and the discussion in Sec. \ref{sec:clasVSquant} for more details. We will, however, see below Thm.\ref{thm:n4nonzerospec} that also in the quantum context ancillary zeros  can be necessary in order to realize a certain spectrum.
	
In the classical context of non-negative matrices, it has been shown in~\cite{JLL96} that if $\Lambda$ is the non-zero part of the spectrum of some $M\in\cM_N(\R_+)$ with ${\rm rank}(M)=r$, then there is an $M'\in\cM_{r^2}(\R_+)$ for which $\spec(M')\backslash\{0\}=\Lambda$. That is, the unboundedness of $N$ in the classical context is related to large or numerous nilpotent Jordan blocks.

	\item \emph{Eq.(\ref{eq:mucond2}) is not necessary.} Note that in the converse part of Thm.\ref{thm:mainIEP} all the imposed conditions on $\Lambda$ but Eq.(\ref{eq:mucond2}) are necessary. While for the analogous classical statement for non-negative matrices in \cite{BoHa91} Eq.(\ref{eq:mucond2}) is evidently  necessary, too, it fails to be so in the quantum context. Let us construct an example of a primitive quantum channel where Eq.(\ref{eq:mucond2}) does not hold:
	
Define the following $d+1$ Kraus operators $K_j\in\cM_d(\C)$:
\be\label{eq:JLLcounterexp} K_j\ :=\ \frac{1}{\sqrt{2}}\;\left\{
\begin{array}{ll}  |j\>\<j+1\;{\rm mod}\; d| & \mbox{ for } j=0,\ldots d-1,\\
 \sum_{k=0}^{d-1}\exp{(\pi i k/d)}\;|k\>\<k| & \mbox{ for } j=d.\end{array}\right.
\ee  The map  $T(\rho):=\sum_{j=0}^d K_j\rho K_j^\dagger$ is then completely positive, trace-preserving and unital.  Moreover, following Eq.(\ref{eq:momentsKraus}) we get $\tr{\hat{T}}> 0$ although $\tr{\hat{T}^2}=0$ if $d>2$. Primitivity of the map $T$ is guaranteed since for $n\geq d$ we have
\be {\rm span}\big\{K_{j_1}\cdots K_{j_n}\big\}\;=\; \cM_d(\C)\ee
which implies primitivity by virtue of \cite{SPWC09}.	

Although $\tr{\hat{T}^2}=0$ in the above example, we note that primitivity generally implies that $\tr{\hat{T}^k}>0$ for all $k\geq p$ where $p$ depends on the dimension only. This follows from Wielandt's inequality which in the classical case of stochastic matrices in $\cM_d(\R_+)$ yields $p=d^2-2d+2$ and in the context of quantum channels on $\cM_d(\C)$ it gives $p=(d^2-2)(d^2-1)$ \cite{SPWC09}.

\item\emph{Classical vs. quantum channels.} Consider a matrix representation $\hat{T}$ of a map $T:\cM_d(\C)\ra\cM_d(\C)$ expressed in terms of matrix units, i.e., $\<k,l|\hat{T}|i,j\>:=\<l|T\big(|i\>\<j|\big)|k\>$ for $i,j,k,l=1,\ldots,d$. Then the $d\times d$ matrix $S$ defined via $S_{ij}:=\<ii|\hat{T}|jj\>$ is entry-wise non-negative if $T$ is a positive map and it is stochastic if in addition $T$ is trace-preserving. That is, every quantum channel $\hat{T}$ contains a stochastic matrix $S$ as a principal submatrix. Eq.(\ref{eq:TfromStoch}) reverses this: it defines a quantum channel from this stochastic submatrix by setting all other matrix elements of $\hat{T}$ to zero.
\end{enumerate}

The close relation between the non-zero spectra of quantum channels and of stochastic matrices implied by Eq.(\ref{eq:TfromStoch}) allows to exploit other results from the classical world. In the following we derive in this way two sufficient conditions for realizable spectra for which a bound on the number of ancillary zeros can be given:
\begin{theorem}[Non-zero spectrum with non-positive real parts]\label{thm:specnonposreals} Let $\Lambda=\{\lambda_1,\ldots,\lambda_n\}$ be a multiset of non-zero complex numbers with $\lambda_1=1$ and ${\rm Re}(\lambda_j)\leq 0$ for all $j\geq 2$, and let $d\in\mathbb{N}$ be the smallest number such that $\mu_1(\Lambda)^2\leq d \mu_2(\Lambda)$. If $\Lambda=\bar{\Lambda}$ and
\be \mu_1(\Lambda)\geq 0\quad \mbox{and}\quad\mu_2(\Lambda)> 0,\label{eq:condforspecposreals}\ee
then there exists a primitive, completely positive and trace-preserving linear map $T:\cM_d(\C)\rightarrow\cM_d(\C)$ such that $\Lambda=\spec(T)\backslash\{0\}$.
\end{theorem}
\proof{ The theorem is a direct consequence of its classical counterpart derived in \cite{LS06} inserted into Eq.(\ref{eq:TfromStoch}). As discussed already in the proof of Thm.\ref{thm:mainIEP} primitivity is carried over from the classical case.}

\begin{theorem}[Real non-zero spectrum for $n=4$]\label{thm:n4nonzerospec} Let $\Lambda=\{\lambda_1,\ldots,\lambda_4\}$ be a multiset of non-zero real numbers. If (i) $\Lambda=\bar{\Lambda}$, (ii) for all $j:|\lambda_j|\leq 1\in\Lambda$ and (iii)  $\sum_{j=1}^4\lambda_j \geq0$, then there exists a completely positive and trace-preserving linear map $T:\cM_4(\C)\rightarrow\cM_4(\C)$ such that $\Lambda=\spec(T)\backslash\{0\}$.
\end{theorem}
\proof{Following \cite{LL78} the stated conditions are sufficient for the existence of a non-negative (and by similarity transformation stochastic) matrix with spectrum $\Lambda$. Using Eq.(\ref{eq:TfromStoch}) again this gives the desired result.}

Thm.\ref{thm:n4nonzerospec} together with Thm.\ref{thm:qubitsolution} shows that ancillary zeros can be necessary to make a certain spectrum realizable. Consider for instance $\Lambda=\{1,1,1,-1\}$. Then Thm.\ref{thm:qubitsolution} implies that this cannot be the spectrum of a quantum channel on $\cM_2(\C)$ whereas Thm.\ref{thm:n4nonzerospec} tells us that it can be the non-zero part of the spectrum of a quantum channel on $\cM_4(\C)$.

Finally, we use the similarity between the classical and the quantum case to characterize the non-zero part of the spectrum of quantum channels with `full Kraus rank'. The latter means that the complex linear span of the Kraus operators of a quantum channel on $\cM_d(\C)$ is the entire matrix space $\cM_d(\C)$. Equivalently, the Choi matrix (or Jamiolkowski state) has full rank. This condition can be seen as the quantum counterpart to entry-wise positivity of stochastic matrices. Note that, in particular, every quantum channel with full Kraus rank is primitive.
\begin{theorem}[Non-zero spectrum under full Kraus rank]\label{thm:nonzerospecfullKraus} Let $\Lambda=\{1,\lambda_2\ldots\lambda_n\}$  be a multiset of non-zero complex numbers. There is a completely positive and trace-preserving linear map $T$ with full Kraus rank such that $\Lambda=\spec(T)\backslash\{0\}$ iff
 $|\lambda_j|<1$ for all $j\geq 2$ and  $\mu_k(\Lambda)>0$ for all $k\in\mathbb{N}$.
\end{theorem}
\proof{ Necessity of the conditions is implied by Eq.(\ref{eq:momentsChoi}) and the general discussion in Sec.\ref{sec:basicprop}. Following \cite{BoHa91} the stated conditions are sufficient for the existence of an entry-wise positive matrix $M$ with $\Lambda=\spec(M)\backslash\{0\}$. As mentioned in the proof of Thm.\ref{thm:mainIEP} this can, by similarity transformation, be made a stochastic matrix $S$ which is again entry-wise positive. Inserting this into Eq.(\ref{eq:TfromStoch}) the latter guarantees full Kraus rank since the Choi matrix of the constructed quantum channel reads
\be \big(T\otimes\id\big)\big(|\Omega\>\<\Omega|\big) = \sum_{ij} S_{ij}|i\>\<i|\otimes|j\>\<j|\geq s\1,\ee
where $s=\min\{S_{ij}\}>0$.
}

\section{Cycles and the peripheral spectrum}\label{sec:peripheralspec0}

We will now have a closer look at the space corresponding to the peripheral spectrum (i.e., eigenvalues which are phases) of a positive linear map $T:\cM_d(\C)\ra\cM_d(\C)$ which is assumed to be either trace-preserving or unital. Denote the complex linear span of all the respective eigenspaces by \be \cX_T:=\mbox{span}\big\{X\in\cM_d(\C)\big|\exists\varphi\in\R: T(X)=e^{i\varphi}X \big\}\label{eq:cyclicspace},\ee
for which we will simply write $\cX$ if the dependence on $T$ is clear from the context. The questions which we address first are: what's the structure of $\cX$? what's the structure of $T$ restricted to $\cX$? and what does this imply for the peripheral spectrum?

Recall that if $T$ is trace-preserving and positive, there cannot be any non-trivial Jordan-block associated to an eigenvalue of modulus one.

The following shows that $\cX$ is a $^*$-algebra (w.r.t. a modified product) and that $T$ acts on it either by permuting blocks or by unitary conjugation on subsystems. The property which we use in order to prove the assertion lies in between positivity and complete positivity: we call a positive trace-preserving map $T$ on $\cM_d(\C)$  a \emph{Schwarz map} if $T^*(A^\dagger A)\geq T^*(A^\dagger)T^*(A)$ holds for all $A\in\cM_d(\C)$. This is in particular true if $T$ is completely positive.\footnote{In fact, $2$-positivity, i.e., positivity of $T\otimes\id_2$, is sufficient but not necessary for $T$ to be a Schwarz map.}

\begin{theorem}[Structure of cycles]\label{thm:cyclestructure} Let $T:\cM_d(\C)\ra\cM_d(\C)$ be a trace-preserving Schwarz map. \begin{enumerate}
\item There exists  a decomposition of the Hilbert space $\C^d=\cH_0\oplus\bigoplus_{k=1}^K\cH_k$ into a direct sum of tensor products\footnote{Note that we may have dim$\cH_0=0$ as well as dim$\cH_{k,i}=1$. Moreover, the direct sum in the decomposition does not necessarily correspond to a block structure in computational basis -- just in \emph{some} basis.} $\cH_k=\cH_{k,1}\otimes\cH_{k,2}$ and positive definite density matrices $\rho_k$ acting on $\cH_{k,2}$ such that
\be \cX_T=0\oplus\bigoplus_{k=1}^K\cM_{d_k}\otimes\rho_k,\label{eq:structureofcX}\ee
where $\cM_{d_k}$ is a full complex matrix algebra on $\cH_{k,1}$ of dimension $d_k:={\rm dim}(\cH_{k,1})$.
That is, for every $X\in\cX_T$ there are $x_k\in\cB(\cH_{k,1})$ such that
\be X=0\oplus\bigoplus_k x_k\otimes\rho_k.\label{eq:Xdecomp}\ee
\item There exist unitaries $U_k\in\cB(\cH_{k,1})$ and a permutation $\pi$, which permutes within subsets of $\{1,\ldots, K\}$ for which the corresponding $\cH_{k,1}$'s have equal dimension, so that for every $X\in\cX_T$ represented as in Eq.(\ref{eq:Xdecomp})
    \be\label{eq:TXpermU} T(X)=0\oplus\bigoplus_{k=1}^K U_k x_{\pi(k)}U_k^\dagger\otimes\rho_k.\ee
\end{enumerate}
\end{theorem}
\proof{
The basic ingredients of the proof are those which already appeared in \cite{WoCi08} where positive maps with unit determinant were characterized.
 Exploiting Dirichlet's lemma on Diophantine approximations\footnote{Dirichlet's lemma states the following~\cite{Schmidt80}: let $x\in\R^m$ and $q>1$ any integer. Then there exist integers $n,p_1,\ldots,p_m$ such that $1\leq n\leq q^m$ and $|x_kn-p_k|\leq 1/q$ for all $k$.} we can find an ascending subsequence  $n_i\in\mathbb{N}$ so that $\lim_{i\ra\infty}T^{n_i}$ converges to a map $I$ with eigenvalues which are either one or zero. Clearly, $I$ is again a trace-preserving, positive linear map which satisfies the Schwarz inequality since all these properties are preserved under concatenation. Moreover, $\cX_T=\cX_I$ is the fixed-point space of $I$. So, statement 1. of the theorem follows from the structure of the space of fixed points provided in \cite{Lind98}.

2. The map $T^{-1}:=\lim_{i\ra\infty}T^{n_i-1}$ is the inverse of $T$ on $\cX$ since by construction $T^{-1}T=I$. Hence, both $T$ and $T^{-1}$ map $\cX\ra\cX$ in a bijective way. The crucial point here is that $T^{-1}$ is a again a positive, trace-preserving map since it is constructed as a limit of such maps. To understand the consequences, consider any pure state in $\cX$, i.e., a density operator $\sigma\in\cX$ which has no non-trivial convex decomposition within $\cX$. Then the image of $\sigma$ under $T$ has to be a pure state as well: assume this is not the case, i.e., $T(\sigma)=\sum_i\lambda_i\sigma_i$ is a non-trivial convex decomposition into states $\sigma_i\in\cX$. Then applying $T^{-1}$ to this equation leads to a contradiction since $\sigma=\sum_i\lambda_i T^{-1}(\sigma_i)$ is not pure. Consequently, both $T$ and $T^{-1}$ map pure states in $\cX$ onto pure states. Note that a pure state can only have support in one of the $K$ blocks, for instance  $\sigma=x\otimes\rho_k$ where $x\in\cB(\cH_{k,1})$ is a rank one projection. Now we know that $T(\sigma)=x'\otimes\rho_{k'}$ for some $k'$ and some rank one projection $x'\in\cB(\cH_{k',1})$. By continuity and the fact that $T$ is a bijective linear map on $\cX$, we have that within $\cX$: (i) every element of  block $k$ is mapped to an element of the block $k'$ (i.e., $k'$ does not depend on $x$), and (ii) the spaces $\cH_{k,1}$ and $\cH_{k',1}$  must have equal dimension. Therefore, there is a permutation $\pi$ which permutes blocks  with equal $d_k$ so that $X\in\cX$ is mapped to
$$T(X)=0\oplus\bigoplus_k T_k(x_{\pi(k)})\oplus\rho_k,$$
with some linear maps $T_k:\cM_{d_k}\ra\cM_{d_k}$. Since the latter have, together with their inverses, to be positive and trace-preserving they must be either matrix transpositions or unitary conjugations~\cite{WoCi08}. Matrix transpositions are, however, ruled out by the requirement that $T$ and thus each $T_k$ is a Schwarz-map.
}

The action of $T$ on $\cX_T$ derived in Eq.(\ref{eq:TXpermU}) now determines the possible peripheral spectra:
\begin{theorem}[Peripheral spectrum]\label{thm:perispec2} Let $T:\cM_d(\C)\ra\cM_d(\C)$ be a trace-preserving Schwarz map. There are integers $n_c,d_c\in\mathbb{N}$ satisfying $\sum_c n_c d_c\leq d$, integers $m_c\in\mathbb{Z}_{n_c}$ and vectors $\mu^{(c)}\in\C^{d_c}$ whose components are phases (i.e., $\forall k:|\mu_k^{(c)}|=1$) so that the peripheral spectrum of $T$ (including algebraic multiplicity) is given by the $\sum_c n_c d_c^2$ numbers
\be \mu_k^{(c)}\bar{\mu}_l^{(c)}e^{\frac{2\pi i m_c}{n_c}}.\ee
Conversely, given such a set of numbers there is a completely positive, trace-preserving and unital linear map $T$ on $\cM_d(C)$ with $d=\sum_c n_c d_c$ so that its peripheral spectrum coincides with the given numbers and it has no other non-zero eigenvalue.
\end{theorem}
\proof{
First note that  the eigenvalue equation $T(X)=\lambda X$ for $X\in\cX_T$ is, via Thm.\ref{thm:cyclestructure}, equivalent to \be \label{eq:reducedeigperi}\forall k:x_{\pi(k)}=\lambda U_k^\dagger x_kU_k.\ee Assume for the moment that $\pi$ is a cycle of length $n_c$, i.e., after relabeling we have $k\in\mathbb{Z}_{n_c}$  and $\pi(k)=k+1\;\mbox{mod}\;n_c$ -- the general case will later reduce to this one by decomposing a general permutation into cycles. All $U_k$'s and $x_k$'s now act on spaces of equal dimension, $d_c$ say. For given $U_k$'s and $\lambda$ the eigenvalue equation determines each $x_k$ from $x_0$ and leads to the additional constraint $x_0=\lambda^{n_c} \tilde{U}^\dagger x_0\tilde{U}$ with $\tilde{U}:=U_0U_1\cdots U_{n_c-1}$. Now define any unitary $U$ on $\C^{d_c}$ for which $U^{n_c}=\tilde{U}$ and consider any of the $d_c^2$ solutions of the eigenvalue equation $y=\mu U^\dagger y U$, $y\in\cB(\C^{d_c})$. Then $x_0= y$ and $\lambda=\mu\exp{[2\pi i m/n_c]}$ leads to a solution of Eq.(\ref{eq:reducedeigperi}) for every $m\in\mathbb{Z}_{n_c}$. By construction the eigenvectors $X$ of $T$ which correspond to different $m$'s will be orthogonal, so that in total the construction leads $n_c d_c^2$ and thus all solutions. If $\mu_1,\ldots,\mu_{d_c}$ are the eigenvalues of $U$, then the multiset of eigenvalues $\lambda$ is given by $\big\{\mu_k\bar{\mu}_j\exp{[2\pi i m/n_c]}\big\}$ with $k,j=1,\ldots,d_c$ and $m\in\mathbb{Z}_{n_c}$.

The  case of a general permutation now follows by decomposing it into cycles (labeled by $c$ and of possibly different lengths $n_c$, which implies that $\sum_c n_c=K$) and observing that the peripheral spectrum of $T$ is just the union of the spectra of all the 'cycle-building-blocks'.

In order to prove the converse we construct a map $T$ guided by the structure appearing in Thm.\ref{thm:cyclestructure} (but now with $\dim\cH_{k,2}=1$). The desired $T$ is obtained as a concatenation of three types of maps: a pinching onto the $\sum_c n_c d_c$ blocks of $\C^d=\bigoplus_c\C^{d_c \otimes n_c}$, a permutation with cycles of length $n_c$ which permute blocks of dimension $d_c$ and, finally, unitary conjugations on each block. In this way we achieve that $T$ is (i) completely positive, unital and trace-preserving, (ii) it projects $\cM_d(\C)$ onto a direct sum of matrix algebras of the form in Eq.(\ref{eq:structureofcX}) (albeit with $\rho_k=1$ and without kernel) and (iii) it acts on this sub-algebra as in Eq.(\ref{eq:TXpermU}). It follows then from the previous analysis that an appropriate choice of the unitary conjugations will lead to the predetermined spectrum.
}

Note that a different converse can be proven along the same lines as in Thm.\ref{thm:perispec2}: for every trace-preserving Schwarz-map $T$ there is a trace-preserving completely positive map $\tilde{T}$ such that $\tilde{T}=T$ on $\cX_T$.\footnote{This is closely related to the well known fact that any  \emph{conditional expectation}, i.e., positive projection onto a *-subalgebra of $\cM_d(\C)$, is automatically completely positive.} That is, complete positivity is not only irrelevant for the peripheral spectrum but for the entire peripheral eigen-system. As shown below Thm.\ref{thm:mainIEP} this does no longer hold for the rest of the spectrum, let alone the corresponding eigen-system.

\section{Classical vs. quantum inverse eigenvalue problems}\label{sec:clasVSquant}

We will briefly review the classical counterpart of our treaties---the inverse eigenvalue problem for stochastic matrices to which we resorted already in Sec.\ref{Sec:momentsANDnonz}. For a more detailed overview see \cite{ChGo05}. Recall that the spectrum of any stochastic matrix is  contained in the unit disc, it contains one, and it is, as a multiset, invariant under complex conjugation. A non-negative matrix\footnote{A matrix is said to be 'non-negative' if all its entries are non-negative.}  is \emph{irreducible} iff the spectral radius is a non-degenerate eigenvalue and the corresponding eigenvector is entrywise positive definite. An irreducible matrix in turn is  \emph{primitive} iff there is only a single eigenvalue whose modulus matches the spectral radius.

 \subsection{The classical inverse eigenvalue problem}
 Every stochastic matrix $S$ is, after a permutation, upper block-triangular with the diagonal blocks being irreducible.  That is, the spectrum of $S$ is the union of spectra of irreducible substochastic matrices. As used in the proof of Thm.\ref{thm:mainIEP} already, every irreducible matrix is, up to a rescaling, similar to a stochastic matrix. For every irreducible matrix $M\in\cM_d(\R_+)$ there is an \emph{index of cyclicity} $k\in\{1,\ldots,d\}$ so that the spectrum of $M$ is invariant under $2\pi/k$-rotations in the complex plane. That is, the eigenvalues come in $k$ rotated $d/k$-tuples. Correspondingly, $M^k$ is block-diagonal with primitive matrices in the diagonal blocks. So finally, the spectrum of any non-negative matrix, stochastic or not, is essentially determined by the spectra of primitive matrices.

 The non-zero part of the spectrum of primitive matrices is completely characterized by Eqs.(\ref{eq:mucond1},\ref{eq:mucond2}) together with the fact that the spectral radius is attained only for a single eigenvalue \cite{BoHa91}. The number of additionally  required zero eigenvalues is, however, generally unbounded. One way to show this is to exploit  the dimension dependent condition
\be \label{eq:JLL} \mu_k(\Lambda)^m\leq d^{m-1}\mu_{km}(\Lambda),\qquad k,m\in\mathbb{N}\ee
which is valid for any non-negative matrix in $\cM_d(\R_+)$ \cite{LL78,Joh81}.
For a multiset $\Lambda$ of three complex numbers Eqs.(\ref{eq:mucond1},\ref{eq:JLL}) are in fact necessary and sufficient for $\Lambda$ being the spectrum of a matrix in $\cM_3(\R_+)$ \cite{LL78}.
For $d=4$, however, these conditions are no longer sufficient.

\subsection{Comparing classical and quantum}
Maybe the most basic difference between the spectra of classical and quantum channels concerns the location of single eigenvalues~\cite{WoPe09}. While for quantum channels on $\cM_d(\C)$ the entire unit disc is accessible for any $d\geq 2$, stochastic matrices in $\cM_d(\R_+)$ obey a dimension dependent constraint on the location of every single eigenvalue: a necessary condition for $\lambda$ to be a valid eigenvalue is that it is contained in the convex hull of all roots of unity of order up to $d$. A complete characterization of the accessible region within the unit disc can be found in \cite{Karp51,DiDy46}.

As a consequence, for certain sequences of four non-zero eigenvalues which obey Eq.(\ref{eq:mucond1}) there is a quantum channel on $\cM_2(\C)$ but no classical channel for any bounded dimension. That is, while on the classical side an unbounded number of zeros have to be added, no extra kernel is required on the quantum side.

As discussed below Thm.\ref{thm:mainIEP} already, it is, however, open whether or not there is a general bound on the number of required zeros in the quantum case. Eq.(\ref{eq:JLL}), which is classically used to bring the dimension into play, fails to hold for spectra of quantum channels: the channel constructed in Eq.(\ref{eq:JLLcounterexp}) leads to $\mu_1>0$ with $\mu_2=0$ and thus violates Eq.(\ref{eq:JLL}) for $k=1,\; m=2$.

Another crucial difference between the classical and quantum case is the breakdown of the problem into irreducible and finally into primitive blocks. The decomposition valid for stochastic matrices  fails to hold for quantum channels and no counterpart of this is known for the quantum case. Maybe the structure of the peripheral eigen-system proven in Thm.\ref{thm:cyclestructure} hints at a more general decomposition.

\section{Applications for time series and correlation functions}\label{sec:appltimecorr}

Suppose we are given a bounded sequence $a\in l^\infty$ which is generated via \be a_t=\< A\big|\hat{T}^t\big|\rho\>,\quad t\in\mathbb{N}_0,\label{eq:linevolution2}\ee from powers of a linear operator $\hat{T}\in\cM_D(\C)$. This may for instance be the discrete time evolution of some expectation value or, as explained later, a two-point correlation function.
Then $a$ contains information about the spectrum of $\hat{T}$. A simple relation between $\spec(T)$ and $a$ is provided by the Cayley-Hamilton theorem, i.e., the fact that $\hat{T}$ is a solution of its characteristic equation. So let $p(x)=\prod_{k=1}^D(\lambda_k-x)=\sum_{k=0}^D c_k x^k$ be the characteristic polynomial of $\hat{T}$, then $\sum_{k=0}^D a_{k+l}c_k=0$ for all $l\in\mathbb{N}_0$.
Depending on degeneracies of the eigenvalues and their Jordan-block structure there can be \emph{annihilating polynomials} of degree smaller than $D$ \cite{Gan74a}.

Conversely, eigenvalues of $\hat{T}$ can be determined as roots of a polynomial which is constructed directly from the sequence $a$. To this end define for each $\tau\in\mathbb{N}_0$ a vector $v_\tau\in l^\infty$ via $v_\tau:=(a_\tau,a_{\tau+1},\ldots)$. If the real linear space $\cV:=span\{v_\tau\}_{\tau\in\mathbb{N}_0}$ has dimension $r<\infty$, then there are coefficients $c_k$ such that for all $l\in\mathbb{N}_0$: $v_{l+r}=\sum_{k=0}^{r-1} c_k v_{l+k}$. These coefficients, which can be determined from $a$, define a polynomial $p(x):=x^r-\sum_{i=0}^{r-1}c_k x^k$. The roots of $p$ then turn out to be eigenvalues of $T$ (since they coincide with the poles of the z-transform discussed below, see \cite{Gan74b} p. 205 ff).

A related connection between $\spec(\hat{T})$ and $a$ is provided by the $z$-transform $\cL:\C\ra\C$:
\be\label{eq:zTrafo} \cL(z):=\frac1z\sum_{t\in\mathbb{N}_0}\frac{a_t}{z^t}\;=\;\<A\big|\big(z\1-\hat{T}\big)^{-1}\big|\rho\>.
\ee
Note that the series which defines $\cL$ converges outside a disc with radius equal to the spectral radius of $\hat{T}$ and $\cL$ is defined inside by analytic continuation.

The poles of $\cL$ correspond to eigenvalues of $\hat{T}$. Depending on the interplay between $\hat{T}$, $A$ and $\rho$, not all eigenvalues may, however, appear as poles.
Consider as an example the sequence
\be a_t= \Big(\mbox{Im}\big[(1-i)^tx^t\big]+y^t\Big)/\sqrt{2}\label{eq:timesexp1},\quad x,y\in[-1,1].\ee
In this case $\cL$ has poles at $z_{1,2}=(1\pm i)x$ and at $z_3=y$. So according to Thm.\ref{thm:qubitsolution} the sequence can be generated by a quantum channel acting on $\cM_2(\C)$ only if $(|z_1|,|z_2|,z_3)\in\cT$.
\begin{figure*}[t]
\centering
\includegraphics[width=0.7\textwidth]{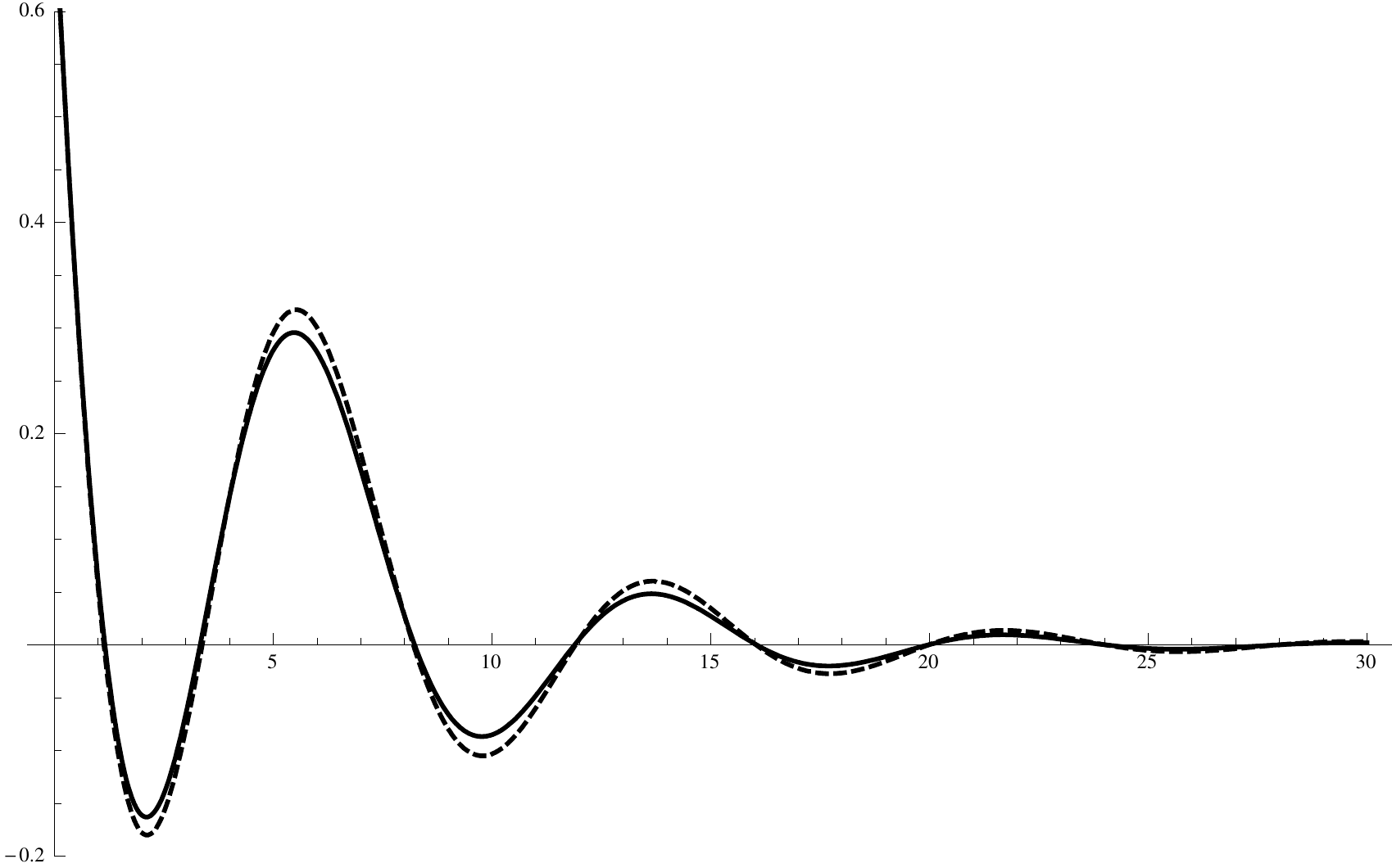}
\caption{Sequences given by Eq.(\ref{eq:timesexp1}) with $y=2/3$. While the solid curve $(x=0.58)$ can be generated by an evolution of a qubit quantum channel, the dashed curve $(x=0.59)$ cannot since the poles of its z-transform do not correspond to valid eigenvalues of a qubit quantum channel. The sequence can, however, be generated by a positive  trace-preserving map. }
\end{figure*}
In fact, it can be generated by a map on $\cM_2(\C)$ with matrix representation
$$\hat{T}=\left(
\begin{array}{cc}
1 & 0 \\ 0 & y
\end{array}
\right)\oplus\left(
\begin{array}{cc}
x & x \\ -x & x
\end{array}
\right),$$
with $\rho$ and $A$ given by $(1,1/\sqrt{2},1/\sqrt{2},0)$ and $(0,1,0,1)$ respectively. So the example can be generated by a qubit channel iff the poles are valid eigenvalues. Also note that $T$ is a positive map iff $|y|\leq 1$ and $|x|\leq 1/\sqrt{2}$.

Using the above relations, knowledge about the spectra of quantum channels can be used to decide whether or not a given sequence $a$ has a representation of the form in Eq.(\ref{eq:linevolution2}) with $T$ being a quantum channel of a certain dimension. Such sequences arise naturally for instance as discrete, homogeneous time-evolutions. They also appear as two-point correlation functions in quantum spin chains whose state is \emph{finitely correlated} \cite{FNW92}. The Kraus operators of the $T$ then correspond to the matrices of the \emph{matrix product state} \cite{PVWC07} and $t+1$ is the distance between the two considered sites on the chain.  So a  detailed spectral analysis might help in this context if one wants to find an accurate finitely correlated approximation to given two-point correlations.

\subparagraph{Acknowledgments} We acknowledge financial support by the Danish research council (FNU), the Spanish grants I-MATH, MTM2008-01366, S2009/ESP-1594 and the European projects QUEVADIS and COQUIT.


\end{document}